\documentclass[twocolumn,noshowpacs,noshowkeys,pra,superscriptaddress,bibliography]{revtex4-1}%
\usepackage{graphicx}
\usepackage{color,pgfplots}
\usepackage{epstopdf}
\usepackage{braket}
\usepackage{amsmath}
\usepackage{caption}
\usepackage{notes2bib}
\usepackage{float}
\usepackage{latexsym}
\usepackage{bbold}
\usetikzlibrary{plotmarks}
\pgfplotsset{compat=1.10}

\usepackage{orcidlink}

\newcommand{\im}{\ensuremath{ \text{i} }}

\begin{document}

\title{Multi-level quantum emitter in an optical waveguide: paradoxes and resolutions}


\author{Ben Lang \orcidlink{0000-0002-5602-9166}}

\begin{abstract}
We theoretically investigate the optical dipole interaction between a multi-level quantum system and a single-mode optical waveguide of any local polarisation. We investigate several paradoxical seeming situations, for example we find a situation in which there exist two non-orthogonal quantum states, each of which results in a photon flux in the opposite direction to the other. We show how, despite appearances, this does not break the unitary requirements of quantum mechanics. We also find that an isotropic quantum emitter can be either reflective or transmissive to light depending on the waveguide polarisation at the emitter location, indeed in the zero loss limit such a system changes from 100\% transmission to 100\% reflection due to an infinitesimal polarisation rotation. An example case for a four level system is also considered, which is found to operate as a non-destructive parity measurement of the photon number.
\end{abstract}


{\let\newpage\relax\maketitle}

\section{Introduction}

An atom (or comparable system, such as a quantum dot or diamond vacancy centre) interacting with a controlled optical mode can be used as a building block to construct a device for processing quantum data. Tight focusing of the light onto the atom, with lenses, cavities or waveguides, is necessary to realise high efficiencies. This tight focusing introduces transverse optical spin angular momentum to the light waves \cite{transverse_long}. This appears as a component of circular polarisation ``rolling'' in a wheel-like way relative to the propagation direction, in contrast to ordinary circular polarisation that spins around the propagation direction like a bullet \cite{photonic_wheels, spin_hall_light}. The rolling breaks the equivalence between the forwards and backwards directions in a waveguide. At a location where the light travelling forward rolls clockwise, that travelling backwards rolls anticlockwise. With the polarisation equivalence between directions broken some atomic transition dipoles may only be able to couple to a single waveguide direction, an effect often described as ``optical chirality'' \cite{chiral_quantum_optics}. For a recent review see \cite{Kok_Reveiw_2025}.


This effect has been studied in a very wide range of physical implementations, surface-plasmons \cite{Unidirectional_surface_waves}, nano-fibres coupled to metallic particles or atoms \cite{Rauschenbeutel, atom_fibre}, whispering gallery resonators \cite{atom_switch} and nano-photonic waveguides hosting quantum dots \cite{lodahl_chiral, Rici_coles1}.

For almost all applications it is desirable to have a strong interaction between the quantum emitter (QE) and the waveguide (WG). For some applications it is useful to have one or more of the dipole transitions within the QE only interact with one of the two waveguide directions. If the dipole describing a relevant transition is circular, then circular polarisation becomes desirable, as with a rolling circular polarisation circular dipole transitions only couple to one of the two WG directions. Circular polarisation and strong QE-WG interactions do not naturally go hand in hand, forcing a compromise between the two \cite{edge_us}. This has lead to several investigations on how to optimise WG structures to achieve the best possible combination of circular polarisation and strong interactions, these have consisted of simple parameter adjustments \cite{optimal_us}, symmetry alterations \cite{better_glide}, the addition of metallic nano particles \cite{Hughes_Metalic_nano_2019} and topological structures \cite{Hughes_Lodahl_desgin, Jalali_Topological_add_drop_23}. An alternative to modifying the WG is to modify the QE. If one has a QE with an elliptical dipole transition, as opposed to a circular one, then complete decoupling of that transition from one WG propagation direction occurs for an elliptical polarization. This means that one operates at a location where the WG polarisation is non-circular, and thus even without optimising the WG the QE-WG interactions are able to be much stronger \cite{Lang_perfect_2022, Rosinski_elliptical_chirality}.

The present paper first presents a fairly general model for calculating the scattering of a photon off, or the emission of a photon from, a multi-level QE interacting with a WG. This builds on our previous work studying two-level systems \cite{Lang_2023}. While the derivations are involved, they produce comparatively simple final expressions, implemented in python \cite{GreenScat_github_repo}. Secondly, this paper uses this model to look at a couple of situations that are of particular theoretical interest. One of these is an apparently paradoxical situation where intuition (and the model) predict that there exist two non-orthogonal quantum states, one of which produces a photon flux only in the forward WG direction, and the other only in the backward WG direction. We find that while this impossible seeming unidirectionality exists, the physics conspires to ensure that one cannot use it to distinguish non-orthogonal states. Another interesting situation considered is that of light scattering from an isotropic QE (one with no preferred dipole). In this case we find that the system becomes highly sensitive to small changes in the polarisation, indeed in the absence of loss modes an infinitesimal change in the WG polarisation at the QE location can be the difference between 100\% reflection and 100\% transmission (the inclusion of loss smooths this out in proportion to the amount of loss).

In section \ref{model_section} we summarise the model while leaving many details to the appendices. The results sections then consider emission and scattering from systems with multiple excited states, and then scattering from a more complicated system with multiple ground states as well as multiple excited states.

\begin{figure}
\includegraphics[scale=0.54]{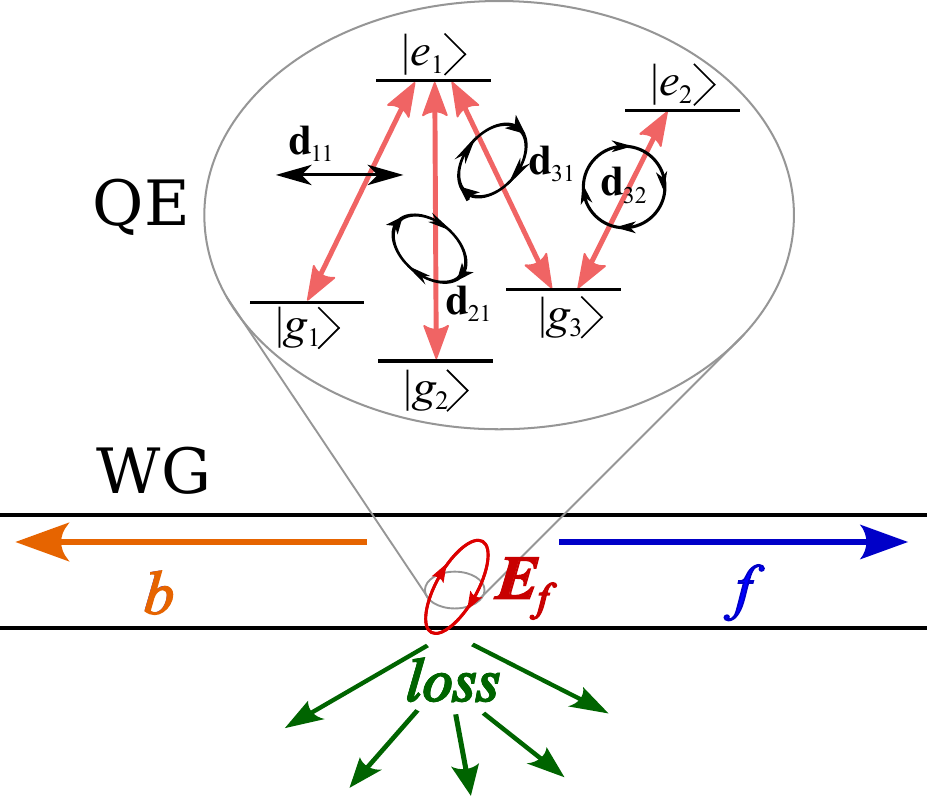}
\caption{Diagram of the interaction between a Quantum Emitter (QE) and Waveguide (WG). The QE is characterized by a set of energy levels with each transition between energy levels controlled by a complex vector transition dipole. The WG possesses a forward and backward optical mode, with the forward mode electric polarisation at the location of the QE controlling the interaction. Loss into non-waveguide modes is also possible.}
\label{scattering_diagram}
\end{figure}

\section{Model}
\label{model_section}

We will first consider the situation where, initially, a single photon is propagating in the forwards direction of the WG and the QE is in one of its ground states. We further assume that the photon is at a single frequency, and thus very long in time. A detailed derivation is provided in appendix \ref{appendix_a}, but some key points will be summarised here. Firstly, the derivation treats the interaction between the QE and WG perturbatively, and finds a pattern that allows all perturbation orders to be collected, meaning there is no approximation introduced by the perturbative approach. However, this pattern relies on the fact that we remain at all times in the single-excitation subspace. This places an important restriction on the types of QE we can consider: they must be \textit{noncascaded}, by which we mean every energy level included must either be a ground state or an excited state, and all allowed transitions must connect a ground state to an excited state (and not, for example, one excited state to another excited state). It also requires the method to use the Rotating Wave Approximation (RWA), which assumes energy conservation. Figure \ref{scattering_diagram} displays an example of an arbitary, compliceted, noncascaded QE.

Interestingly, the resulting equation turns out to have exactly the same form as that for a two-level QE \cite{Lang_perfect_2022, Lang_2023}, except with various terms expanded from complex numbers into matrices and vectors.

Specifically, for a two-level QE with only one transition the final (long time) complex amplitude of finding the photon in the state $\ket{1_m}$ is given by\cite{Lang_2023}:

\begin{equation}
\gamma_{m} = \delta_{f}^{m} - \frac{ \mathbf{E}_{m}^*(\mathbf{r}) \cdot \mathbf{d} \,\,\,
\mathbf{d}^{*} \cdot  \mathbf{E}_{f}(\mathbf{r})}{N(\Gamma + \im \Delta)},
\label{scattering_main}
\end{equation}
\noindent
With $\delta$ a Kronecker delta function taking the value 1 if the mode under consideration (mode $m$) is the forward mode (where the photon was initially), and zero otherwise. Reading the equation from right to left, first the input photon uses the electric field of the input mode ($\mathbf{E}_f$), evaluated at the QE location ($\mathbf{r}$) to interact with the transition dipole ($\mathbf{d}$) with an interaction proportional to the field-dipole inner product. Similarly, the excited state then decays, producing interaction with the mode $\mathbf{E}_m$ in proportion to another dot product. In this second case the causality is from dipole to electric field, resulting in the conjugate being the other way around. (Cause is unconjugated, effect is conjugated.)

In the denominator we see $\Gamma$, which is the total excited state lifetime, and $\Delta$, the detuning between the photon input frequency and the transition frequency of the two-level system. $N$ is a normalization factor, which depends on how the electric fields $\mathbf{E}_f$ and $\mathbf{E}_m$ are normalized. In principle it is always possible roll this into the normalization of the electric fields and set $N=1$ (although this requires an unusual field normalisation).

In a more complicated system with multiple ground states and excited states the basic form of this equation remains unchanged, but the individual terms gain in dimensionality, becoming vectors and matrices. If the multi-level system  starts in ground state $\ket{g_r}$ with a single photon traveling forwards in the WG towards the QE, $\ket{1_f}$, the final (long time) complex amplitude in the state $\ket{1_m, g_k}$ is given by

\begin{equation}
\gamma_{m,k} = \delta_{f,r}^{m,k} -  \mathbf{E}_{m}^*(\mathbf{r}) \cdot \overrightarrow{\mathbf{d}_{k:}} \, \left(N(\overline{\Gamma} + \im \overline{\Delta}) \right)^{-1}\,
\overrightarrow{\mathbf{d}_{r:}^{*}}^{\text{T}} \cdot  \mathbf{E}_{f}(\mathbf{r}),
\label{scattering_main}
\end{equation}
\noindent
where $\overrightarrow{\mathbf{d}_{k:}} =  \begin{pmatrix} \mathbf{d}_{k1}, & \mathbf{d}_{k2}, &\cdots & \mathbf{d}_{km} \end{pmatrix}$ with $\mathbf{d}_{nm}$ the transition dipole connecting ground state $n$ to excited state $m$, and $\delta_{f,r}^{m,k}$ takes value of zero, except when $f=m$ and $r=k$ where it is $1$. Reading equation (\ref{scattering_main}) from right to left we first see the photon in the forwards mode use the electric field of this mode evaluated at the location of the QE ($\mathbf{E}_f(\mathbf{r})$) to interact with the QE. From the initial ground state, $\ket{g_r}$ there are a set of transitions available (one to each excited state), which fill the vector $\overrightarrow{\mathbf{d}_{r:}^{*}}^{\text{T}}$, which has length equal to the number of excited states, where each element of this vector (a particular $\mathbf{d}$) is a vector in real space. The interaction strengths of each of these transitions are proportional to the dot products between the electric field and transition dipole (the dot products indicated by $\cdot$ apply on the inner vectors, the multiplications implicit in adjacent terms apply at the outer level). Skipping over the $\left(N(\overline{\Gamma} + \im \overline{\Delta}) \right)^{-1}$ for now, we see that we then calculate further $\mathbf{E}^*\cdot\mathbf{d}$ products to see how much each excited state generates amplitude in the specific choice of mode and ground state we are considering.

These $\mathbf{E}^*\cdot\mathbf{d}$ bookends are intuitive and match what is seen in the case of a two level system.

In the two-level case there was a denominator. For multilevel systems, we instead have a matrix inverse term, $\left(N(\overline{\Gamma} + \im \overline{\Delta}) \right)^{-1}$ where both $\overline{\Gamma}$ and $\overline{\Delta}$ are square matrices of size equal to the number of excited states. $\overline{\Delta}$ is quite simple, it is a diagonal matrix with each element given by the frequency detuning of an excited state from the input state. Similarly, the diagonal elements of $\overline{\Gamma}$ are just the decay constants of each excited state. The off-diagonals of $\overline{\Gamma}$ represent the fact that sometimes a superposition of two excited states can decay at a rate different from either of the two separately, due to an interference between the outgoing radiation \cite{Bea_2020_array}. They take the same mathematical form as the equations for superradiance/subradiance between separate atoms, as detailed in appendix \ref{appendix_Greens_Hamiltonian}. As for the two-level case, $N$ is a normalization factor that can be rolled into the electric fields if desired.

Instead of throwing a photon at the QE, it is alternatively possible to model the behavior of the QE-WG system after the QE is initialized in the excited state manifold. From this start point population can move around within the excited state manifold (from one excited state to another) or it can move to the ground state manifold along with photon emission, which is irreversible as we assume the photon doesn't come back. A general expression including both possibilities is given by

\begin{equation}
\begin{split}
\frac{d\hat{\rho}_a(t)}{dt} =& \sum_{xy} \frac{\im}{\hbar} \left( \overline{\Gamma}_{xy} \ket{e_x}\!\bra{e_y} \hat{\rho}_a - \overline{\Gamma}^*_{xy} \hat{\rho}_a \ket{e_x}\!\bra{e_y}
\right) \\
&- \frac{\im}{\hbar} [\hat{H_0}, \rho_a(t)] \\
&-\frac{\im}{\hbar} \sum_{xynm} \overline{V}_{xynm} \ket{g_n}\!\bra{e_x} \rho_a \ket{e_y}\!\bra{g_m} \,,
\end{split}
\label{Emission_Equation}
\end{equation}
with $\rho_a$ the density matrix of the QE after the optical modes are traced out. The terms related to $\overline{\Gamma}$ induce population to move between excited states while the array of values in $\overline{V}$ control the irreversible decay to ground states. As shown in the appendix the elements of these matrices can be calculated as follows:

\begin{equation}
\overline{V}_{xynm} = \overline{W}_{xynm} -  \overline{W}_{yxmn}^{*}
\end{equation}
\begin{equation}
\overline{\Gamma}_{xy} = \sum_n\overline{W}_{xynn} 
\end{equation}
(note the reversed and repeated subscripts on some $\overline{W}$s - these are needed), with
\begin{equation}
\overline{W}_{xynm} = - \,\frac{\mathbf{d}_{nx} \cdot \textbf{G}^{*}(\mathbf{r}, \mathbf{r}, \omega_n) \cdot \mathbf{d}_{my}^{*}}{\epsilon_0}\,.
\end{equation}
Here, $\textbf{G}$ is the electromagnetic Green's function. Green's functions connect two points in space, in this case the repeated $\mathbf{r}$ argument is the location of the QE, meaning that it is the part of the Green's function connecting this point back to itself that is relevant. $\epsilon_0$ is the permittivity of free space. The Hamiltonian, $\hat{H}_0$ is the noninteracting Hamiltonian of the QE, which is simply given by the energies of all the QE emitter states written to fill the elements of a diagonal matrix, or $\hat{H}_0 = \sum_n E_{gn} \ket{g_n}\bra{g_n} + \sum_m E_{em} \ket{e_m}\bra{e_m}$.

The Green's function itself strongly depends on the waveguide mode electric field \cite{Hughes_laser_review, stephen_beta}::

\begin{equation}
\textbf{G}(\mathbf{r},\mathbf{r}, \omega) = \frac{\im a \omega}{4|v_g|}\left( \mathbf{E}_f \mathbf{E}_f^* + \mathbf{E}_b \mathbf{E}_b^* \right) + \textbf{G}_\text{loss}
\label{Greens_function}
\end{equation}
where $a$ is the waveguide periodicity, $v_g$ the group velocity of the waveguide modes and $\textbf{G}_\text{loss}$ a term controlling non-waveguide contributions, for example from loss modes. This Green's function is valid for waveguide mode electric fields given for modes equally spaced in $k$-vector and with $\int \mathbf{E}_f(\mathbf{r})\mathbf{E}_f^*(\mathbf{r})\epsilon(\mathbf{r}) d^3\mathbf{r} = 1$. Different normalisation or indexing conventions will give different prefactors. Using this convention the normalisation variable arising in the scattering equations $N = 2|v_g| \epsilon_0 / (\im a \omega)$.

The backward waveguide mode is given by the time reverse of the forward one, meaning that $\mathbf{E}_b = \mathbf{E}_f^*$. As conjugation maps a left circular field to the orthogonal right circular this makes polarisation a highly important parameter. For linear polarisations, the coupling rate of any dipole will be equal to the forwards and backwards modes ($\mathbf{E}_b \cdot \mathbf{d}^* = \mathbf{E}_f \cdot \mathbf{d}^*)$, but in general the two rates can be unequal. Both circular and linear polarisation are highly non-generic, representing the extreme possibilities of a generally elliptical polarisation. Similarly, the transition dipoles are generically elliptical, with linear and circular as special cases.

\section{Results}

We will use simple units throughout. The waveguide period is $a=1$, the waveguided light's group velocity $v_g = c/10$, $|\mathbf{d}| = 1$ (for all dipoles) and $|\mathbf{E_f}|=1$. We also take $\omega = 1$, $\hbar =1$ and $c=1$. Consequently, a simple two level QE with transition dipole $\mathbf{d}$ at a WG location with polarisation $\mathbf{E}_f$ prepared in the excited state decays via emission of a photon in the forward direction with rate constant of $|\mathbf{d}^* \cdot \mathbf{E}_f|^2 / (2 v_g) \leq 5$ inverse time units. This bound is saturated when $\mathbf{d} = \mathbf{E}_f$.

\subsection{Distinguishing non-orthogonal Quantum states}
\label{Nonorthogonal_dipoles}

The coupling to the forwards mode by a dipole is proportional to $|\mathbf{d}^*\cdot \mathbf{E}_f|^2$, while for the backwards mode it is proportional to $|\mathbf{d}^*\cdot \mathbf{E}_b|^2 = |\mathbf{d}^*\cdot \mathbf{E}^*_f|^2$. This leads to the somewhat surprising observation that at a point of elliptical polarisation the choice of dipole that radiates entirely forwards ($\mathbf{d}_1^* \cdot \mathbf{E}_b = 0$) is \emph{not orthogonal} to the one that radiates entirely backwards ($\mathbf{d}_2^* \cdot \mathbf{E}_f = 0$). A quantum state containing a photon travelling forward is orthogonal to one with a photon travelling backwards, such that (neglecting any loss channels) the two non-orthogonal dipoles produce fully orthogonal states of light. This may sound paradoxical, but it is important to recall that dipoles are not quantum states, but instead complex vectors in real space. The orthogonality of dipoles is a fact of geometry in real 3D space as measured in meters. When it is said two quantum states are orthogonal, this is a statement about Hilbert space. So, thus far, there is no paradox, so long as one is careful to keep the two different uses of the word ``orthogonal" clearly separated.

As just stated, dipoles are not quantum states, but vectors in real space. Furthermore, these real-space dipole vectors are not in general directly related to quantum states, but instead appear as terms inside \textit{operators}, such as in equ.\ref{Rotating_Hamiltonian}. In this sense transition dipoles have more in common with polarisers, than with polarisations.

However, there is one situation in which we can establish a direct mapping between quantum states and effective dipole vectors. Consider a 3-level system in an \emph{V} arrangement with the excited states degenerate and the transitions possessing orthogonal dipoles of equal magnitude. Choosing a superposition of the two excited states as $\ket{\psi} = \alpha \ket{e_1} + \beta \ket{e_2}$ one can produce any \emph{effective} dipole describing the overall decay to the ground state by appropriate choice of $\alpha$ and $\beta$, as the effective radiative dipole of the overall decay $\ket{\psi} \rightarrow \ket{g_1}$ is given by:

\begin{equation}
\mathbf{d}_{\text{Eff}} = \alpha \mathbf{d}_{11} + \beta \mathbf{d}_{12},
\label{effective_dipole}
\end{equation}

for $\mathbf{d}_{11,12}$ the dipoles of the two arms of the \emph{V} system.

Consider the local waveguide polarisation $\mathbf{E}_f = \begin{pmatrix} 2 & i \end{pmatrix}$ and dipoles $\mathbf{d}_{11} = \begin{pmatrix} 1 & 0 \end{pmatrix}$ and $\mathbf{d}_{12} = \begin{pmatrix} 0 & 1 \end{pmatrix}$. With these, the state  $\ket{\psi_f} = (i \ket{e_1} + 2 \ket{e_2})/\sqrt{5}$ only radiates in the forwards direction, while $\ket{\psi_b} = (-i \ket{e_1} + 2 \ket{e_2})/\sqrt{5}$ only in the backwards direction. But these states have a $9/25$ mod-squared overlap! We appear to have a situation where non-orthogonal quantum states of the QE produce orthogonal states of light - clearly a violation of quantum theory.

\begin{figure}
\includegraphics[scale=0.62]{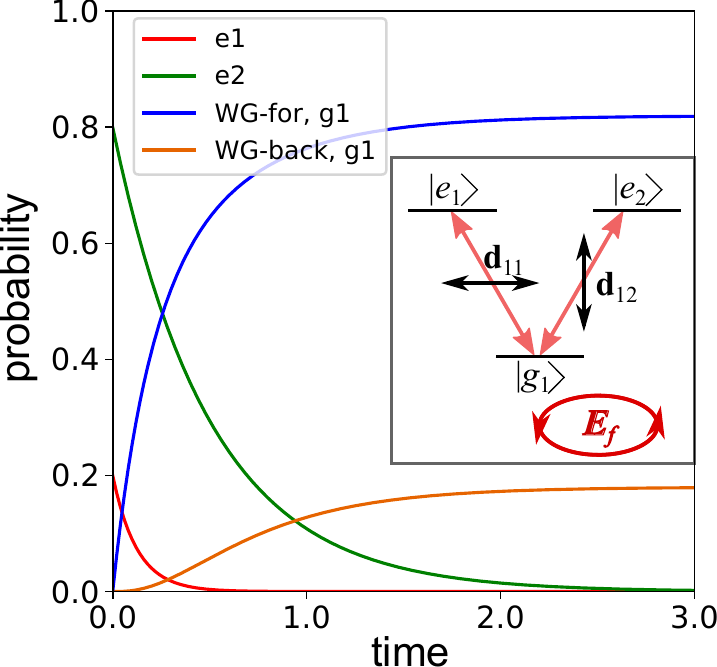}
\caption{Time evolution of QE-WG system. Inset: picture of a \emph{V}-like system with degenerate excited states and orthogonal linear transition dipoles. The WG has local forward mode has polarisation $E_f = (2, \im)/\sqrt{5}$}.
\label{Ellipse_paradox}
\end{figure}

To test this prediction, the emission model of equ.(\ref{Emission_Equation}) is time integrated with these parameters using the QuTiP tool \cite{QuTiP}, as shown in fig.\ref{Ellipse_paradox}.

The figure shows the probability of the system occupying each of four states over time. At $t=0$ we see $\bra{e_1} \rho(t) \ket{e_1} = 0.2$ and $\bra{e_2} \rho(t) \ket{e_2} = 0.8$ as expected for $\ket{\psi_f}$. We also see that, at \textit{short} times, the probability of a photon existing in the forward mode rapidly increases, while the probability of a photon existing in the backward mode remains zero. So, at these short times, the radiation is unidirectional as predicted.

However, as time advances, notice that the probability of the system being found in $\ket{e_1}$ decays very rapidly, much faster than the decay of $\ket{e_2}$. The reason for this can be read off the inset diagram: the dipole $\mathbf{d}_{11}$ has a much stronger overlap with the WG polarisation that $\mathbf{d}_{12}$. Indeed, $|\mathbf{E}_f \cdot \mathbf{d}_{11}|^2 = 4 \,|\mathbf{E}_f \cdot \mathbf{d}_{12}|^2$, giving a four-times faster decay of the population from $\ket{e_1}$ than $\ket{e_2}$. This means that at later times the careful ratio of the complex amplitudes on the two excited states needed to maintain unidirectional behaviour does not last. Thus, as time progresses we start to see the probability of a photon being found in the backward WG mode increase. 

So, with what level of fidelity can we differentiate the two initial states $\ket{\psi_f}$ and $\ket{\psi_b}$? We know $|\braket{\psi_f|\psi_b}|^2 = 9/25$, meaning that the distinguishability of these states, as measured by the trace-distance, is $\sqrt{1 - 9/25} = 4/5$. We also see from the model that in the long term $\ket{\psi_f}$ gives a 0.8 chance of the photon going forward. With $\ket{\psi_b}$ things are symmetrical, with a 0.8 chance of the photon going backward at long times, consistent with this distinguishability of 4/5.

Assuming that we can measure both the direction of the radiated photon, and the emission time, can we do better? (Physics requires that the answer is no, but we will check to make sure.) Sometimes, the photon will be detected at very short times and thus its direction perfectly indicates the initial state. However, sometimes the photon will not be detected until later when the emission is less directional and less is learned. So, the overall distinguishability involves multiplying the intensity of the emission with its directionality and integrating over time. However, the total probability (at long times) of the photon being in the forward or backward modes is given by this same integral, and so the time resolved measurement cannot, in expectation, do better than the 4/5s probability already discovered.

To reiterate, we find that it is indeed true, despite sounding paradoxical, that there exist two non-orthogonal quantum states, one of which produces a photon flux in only the forward direction, and the other only the backward direction. However, a photon flux is not a photon, these are only the \textit{instantaneous} fluxes at a specific moment in time, and when the system is allowed to evolve freely these fluxes evolve such that the overall probability of a photon being radiated either forwards or backwards exactly preserves the non-orthogonality of the initial states.

\subsection{Basis Symmetry}

Given a \emph{V} type system with degenerate excited states and equal magnitude, orthogonal dipoles (as in the last subsection), equ.(\ref{effective_dipole}) gives rise to an interesting symmetry. This symmetry is that the excited state manifold can be rotated to any basis, with the dipoles undergoing an analogous rotation

\begin{equation}
\begin{pmatrix} \ket{e_a} \\ \ket{e_b} \end{pmatrix} = \overline{U} \begin{pmatrix} \ket{e_1} \\ \ket{e_2} \end{pmatrix} \,,
\end{equation}
\begin{equation}
\begin{pmatrix} \mathbf{d}_{1a} \\ \mathbf{d}_{1b} \end{pmatrix} = \overline{U} \begin{pmatrix} \mathbf{d}_{11}  \\ \mathbf{d}_{12}  \end{pmatrix} \,,
\end{equation}
with $\overline{U}$ a unitary matrix.

This means that, the system depicted in the fig.\ref{Ellipse_paradox} could equally have been drawn with alternative dipoles on the transitions, for example circles of opposite helicity, this change being simply a shift in basis.

\subsection{Scattering from an Isotopically polarisable QE}

This highlights how the special case of a $\emph{V}$ like system with degenerate energies and equal-magnitude orthogonal dipoles is a highly symmetric system. Within the 2D basis spanned by the dipoles this system is isotropically polarizable. A similar setup with three degenerate excited states with mutually orthogonal dipoles would be isotropically polarisable in three dimensions - however the electric field of the waveguide mode at the QE location, $\mathbf{E}_f( \mathbf{r})$, would at most span two of those dimensions making the third redundant.

When a two-level $\emph{I}$ type systems interacts with an input photon it can either reflect or transmit the photon, while applying a phase to both transmitted and reflected light that depends critically on the WG polarization and the dipole of the transition. These phase relations posses a rich geometrical and topological structure \cite{Lang_2023}. The isotropically polarizable system is interesting, as it erases the dipole vector - this QE possesses both orthogonal dipoles equally strongly.

In the previous section we neglected interactions with non-WG modes, as weak loss channels didn't qualitatively impact the conclusions. However, in this scattering situation non-WG mode interactions take on an essential role, and must be considered. For simplicity we make the non-WG mode interaction independent of dipole, which implies two equally significant loss modes with polarisations orthogonal at the QE location. We consider two levels of loss, first one where $\mathbf{d}\cdot \mathbf{G}_{\text{loss}}\cdot \mathbf{d}^* = 0.2$ (independent of $\mathbf{d}$), and a weaker level of loss where it is instead $0.003$. These values should be compared to the WG coupling strength, for which we assume $v_g =c/10$ and thus $\mathbf{d}\cdot \mathbf{G}_{\text{WG}}\cdot \mathbf{d}^* \leq 1/v_g = 10$ (where we only say $\leq$ as it depends on the WG polarisation and dipole orientation). For a linear dipole perfectly matched to a linear WG polarisation our parameters and the loss rate of 0.2 (0.003) give a total probability that an initially excited QE puts the radiated photon into a WG mode ($\beta$-factor) of $10 / 10.2 \approx 98\%$ ($10 / 10.003 \approx 99.97\%$).

\begin{figure}
\includegraphics[scale=0.7]{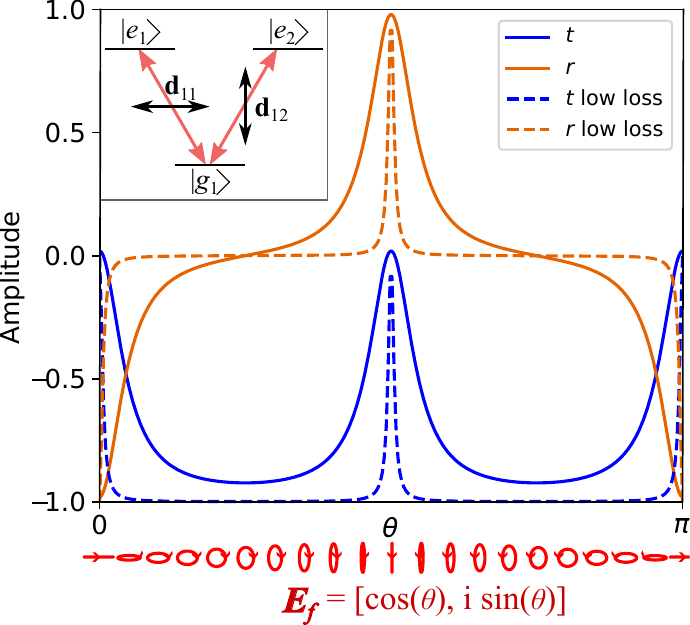}
\caption{Photon scattering from an isotropic QE system, represented by \emph{V} type energy levels as shown in the inset. The transmission ($t$) and reflection ($r$) coefficients are plotted for two different levels of non-WG mode loss mode coupling (see text), as a function of the polarisation of the WG at the QE location, depicted along the $x$-axis.}
\label{isotropic_scattering}
\end{figure}

Using the scattering model, a photon is inserted in the forward direction and the long-term amplitudes of $\ket{1_f}$ (transmission, $t$) and $\ket{1_b}$ (reflection, $r$) are calculated, as a function of $\mathbf{E}_f$, parameterized as $\mathbf{E}_f = (\cos(\theta), \im \sin(\theta))$ as $\theta$ is swept from $0$ to $\pi$. In many types of optical waveguide, such as photonic crystal waveguides, all these polarisations exist somewhere in the waveguide mode, and which is relevant it determined by which occurs at the location of the QE \cite{edge_us}. The results are shown in fig.\ref{isotropic_scattering}. 

First, notice that for the two circular polarisations ($\theta = \pi/4$ and $3\pi/4$) the reflection coefficient is zero, while the transmission coefficient is close to $-1$, indicating transmission of the photon with the QE applying a phase shift (the phase in transmission is measured relative to the transmitted phase if no QE existed). The QE scatters some amplitude out of the waveguide altogether, hence $|t|<1$. These responses for circular polarisations are not surprising - a basis can be adopted in which the two transition dipoles of the $\emph{V}$ QE are oppositely circular. When $\mathbf{E}_f$ is circular it will only couple to one of these two arms of the system, and it is well known that a two level system with circular dipole matching a circular $\mathbf{E}_f$ allows photons to be transmitted with a phase shift \cite{Young_PRL}.

However, this same basis selection trick \emph{cannot} be used across the whole range. For an arbitrary elliptical $\mathbf{E}_f$ things are more complicated. We know that the transition dipole orthogonal to this $\mathbf{E}_f$ does not get excited \emph{directly} by the input photon. Thus we can adopt a basis ($\mathbf{d}_{1a}$, $\mathbf{d}_{1b}$) such that $\mathbf{d}_{1b}^* \cdot \mathbf{E}_f = 0$. Thus, the \emph{initial} interaction with the input photon excites the system towards the excited state $\ket{e_a} = \cos(\theta) \ket{e_1} + \im \sin(\theta) \ket{e_2}$ with associated dipole $\mathbf{d}_{1a} = \mathbf{E}_f$. However, this is not an equilibrium response, because in general the states $\ket{e_1}$ and $\ket{e_2}$ will decay at different rates, as seen in section \ref{Nonorthogonal_dipoles}. This means that, even if $\mathbf{d}_{1a} = \mathbf{E}_f = \cos(\theta) \ket{e_1} + \im \sin(\theta) \ket{e_2}$ were generated, the differing decay rates of $\ket{e_1}$ and $\ket{e_2}$ would see the effective dipole evolve away from this configuration. In practice, throughout the scattering process the state of the QE is a compromise between how it is ``being pushed" (excitation from the input photon) and its tendency to bounce back (decay). So while the input photon pushes the system towards the state $\ket{e_a}$, decay processes simultaneously pull it in a different direction by decaying the $\ket{e_1}$ and $\ket{e_2}$ components at different rates. 

Interestingly, neglecting loss modes, the overall effect is to recreate the effect of a two-level system with the dipole $\mathbf{d} = (\sin(\theta), \im \cos(\theta))$, that is, a dipole that is 90\textdegree rotated from $\mathbf{E}_f$. As we discussed in a previous paper \cite{Lang_2023}, this dipole is special because, it decouples from the backward waveguide mode, ensuring a reflection coefficient of zero. Indeed, in the figure we see that for the low loss curve $r\approx0$ across most of the range. For exactly linear polarisation this logic does not work, as $\mathbf{d}_{12}$ decouples from both the forwards and backwards modes, so the $\emph{V}$ can be modelled as a linear dipole two level system, for which we expect photon reflection \cite{Lang_2023}. Hence, the idealized (no loss mode) situation displays a discontinuity, with $t=-1$, $r=0$ for any polarisation that is not exactly linear, and $t=0$, $|r|=1$ for exactly linear polarisation. This is very surprising, as it implies that an infinitesimally small change in the polarisation at the QE location can move us from 100\% reflection to 100\% transmission. Given an appropriate waveguide this might correspond to an infinitesimal change in the QE position.

The cause of this step discontinuity is as follows. As $\mathbf{E}_f$ gets closer and closer to $\mathbf{E}_f = (1, 0)$ the generation rate of $\ket{e_2}$ falls, but the lifetime of the $\ket{e_2}$ population that is generated grows, with the result that overall the equilibrium contribution from $\ket{e_2}$ grows as the polarisation gets closer to $(1, 0)$. However, this trend abruptly ceases to apply when the polarisation is exactly equal to $(1,0)$, as here the generation rate is zero, and the infinite lifetime is irrelevant as no population is generated in the first place.

This is why coupling to non-WG lossy modes becomes so important. When these loss modes are included, even at $\mathbf{E}_f = (1, 0)$ the state $\ket{e_2}$ has a finite lifetime. The loss channels smooth out the discontinuity, and, as seen in fig.\ref{isotropic_scattering} higher loss results in more smoothing.

There is, to our knowledge, no physical law preventing a sufficiently carefully fabricated WG from getting arbitrarily small losses, or equivalently a $\beta$-factor arbitrarily close to 1. This could be done for example by containing the WG in a large 3D photonic crystal with a full bandgap \cite{Yablonovitch_full_bandgap_1987}. However, as loss is reduced, the sensitivity to the WG polarisation at the QE location becomes ever higher, suggesting the possibility of an arbitrarily precise measure of the QE position. Its not clear how Heisenberg uncertainty would be maintained in this case, as there seems to be no mechanism that to increase the spread in the QE momentum.

\section{Multiple Ground States}


Thus far we have considered a single ground state coupled to multiple excited states. The opposite case, multiple ground states sharing a single excited one, is interesting because it allows the QE to retain information in the ground state manifold, potentially making some parameter regimes useful for quantum memories. The simplest system with two ground states is three levels in a $\Lambda$ configuration. With this energy level arrangement, if each of the two dipoles couples to only one WG direction, then a fascinating switch-like physics emerges as described in \cite{atom_switch}. 

We consider the next simplest case, four levels, two excited states and two ground states, giving four transitions in total in a \emph{IXI} configuration.

If either of the two excited states is excluded (for example by being far detuned or having weak dipole transitions) then it starts to behave more and more like a three level $\Lambda$ system. Consequently, we will focus on the situation where the excited states are near degenerate and contribute equally.

In a charged quantum dot under the influence of an in-plane magnetic field (called Voigt geometry) the transition dipoles are as shown in the figure, with each ground state coupling to one of the two excited states via a horizontal dipole, and the other via a vertical dipole that is $\pi/2$ out of phase relative to the horizontal dipole, IE $\mathbf{d}_{11} = \mathbf{d}_{22} = (1, 0)$, $\mathbf{d}_{21} = \mathbf{d}_{12} = (0, \im)$ \cite{Steel_Voigt_2013}.

\begin{figure}
\includegraphics[scale=0.65]{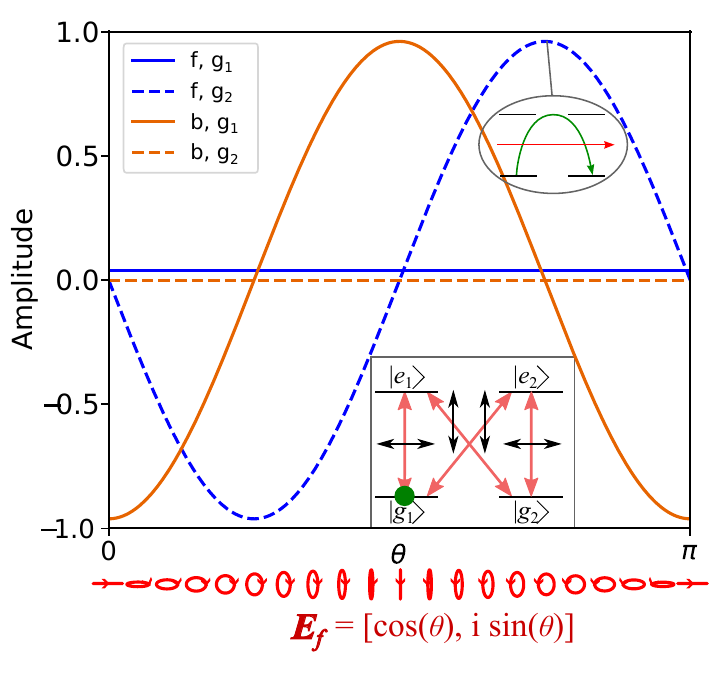}
\caption{Photon scattering from a four-level QE as drawn in the inset, as a function of the polarisation of the WG at the QE location, depicted along the $x$-axis. For circular polarisation the photon transmits while toggling the ground state of the QE, as shown for one of the two circular polarisations in the bubble.  Green circle on insets: initial state.}
\label{IXI_fig}
\end{figure}

If fig.\ref{IXI_fig} the scattering coefficients are plotted for this QE placed in WG locations with various local polarisations, using the same parametrisation as used previously in fig.\ref{isotropic_scattering}. Losses are kept fixed at $\mathbf{d}\cdot \mathbf{G}_{\text{loss}}\cdot \mathbf{d}^* = 0.2$. The QE is initialised in $\ket{g_1}$, the probability amplitudes of $\ket{1_f, g_1}, \ket{1_f, g_2}, \ket{1_b, g_1}, \ket{1_b, g_2}$ after a single forward going photon is inserted, are plotted. The mod-squares of these four curves do not quite sum to 1 due to the loss channel.

First, consider a linear WG polarisation. In this case the photon is reflected, and the QE remains in the first ground state. This occurs equally for any linear polarisation, not just the horizontal and vertical polarisations occurring in the plot. For linear polarisations the \emph{IXI} QE acts essentially as a reflector.

In contrast, for a circular polarisation something more interesting happens. In this case the photon is transmitted, but the QE flips to the other ground state. For one of the two circular polarisations this is accompanied by a phase shift. If the QE were instead initialised in the other ground state, or the photon were input from the other side, then the photon would still transmit while flipping the ground state of the QE.

If multiple photons were injected one at a time the QE would flip ground state in response to each of them in turn, meaning that the final ground state occupied would indicate the parity (even-ness or odd-ness) of the number of photons to have passed. As the QE is equally sensitive to photons travelling in both of the two WG directions this is a two-mode parity measurement, indicating if the total photon number in the two modes (forward and backward) when added together is even or odd. Assuming the ground state before the photons are incident was known, and the final ground state was measured, this would constitute a non-destructive two-mode parity measurement. The inevitable existence of loss modes in any real system means that the transmission probability of each photon is not unity, it is instead given by $\beta^2$ for $\beta$ the QE-WG $\beta$-factor. If the QE is placed in an excited state, then the $\beta$-factor is the probability that it decays by photon emission into a WG mode. In real photonic crystal waveguide devices $\beta$-factors of 99\% are achievable \cite{Lorenzo_2019_beta, high_beta_measurement}.

When a photon escapes due to loss the measurement has not successfully been non-destructive (as the photon number is changed). The total failure probability is $1 - \beta^{2N}$ for $N$ the number of photons. Two-mode photon parity measurements are in some cases very useful \cite{Taehyn_two_mode_parity_2025}, meaning that this parameter configuration may be worthy of further study.

Note that the factors of $\im$ on the diagonal transition dipoles are relevant. If instead of $\mathbf{d}_{21} = \mathbf{d}_{12} = (0, \im)$ we set $\mathbf{d}_{21} = \mathbf{d}_{12} = (0, 1)$ then the system behaviour is somewhat different. For example, with these dipoles at a point of circular polarisation the QE reflects photons incident from either side, while flipping ground state each time it reflects a photon.

\section{Conclusion}

We investigated the interaction of a multilevel quantum emitter (QE) with an optical waveguide (WG). A reasonably general model, combining and building on existing results, was presented. This model was used to investigate a few particularly interesting scenarios. In the first, we find a situation in which two non-orthogonal quantum states exist, each of which implies a photon flux in the opposite direction to the other. Such a situation appears paradoxical, as non-orthogonal quantum states should not be fully distinguishable. We showed that, the paradoxical behaviour is real, but persists for only a short time, exactly short enough to achieve the proper non-orthogonality of the final (long time) optical states. Secondly, we considered the case of photon scattering from an isotropically polariseable dipole. In this case, we found that for an idealised situation where losses are zero the photon is transmitted for all WG polarisations, except those that are exactly linear, for which the photon is reflected. Giving rise to an, in-principle, infinite sensitivity to an infinitesimal change in the polarisation away from linear. The inclusion of loss modes smooths things out, meaning that (with loss) even polarisations close to linear generate some reflectivity.

Neither of these cases were selected for realism, but instead because they are highly symmetric situations of correspondingly high theoretical interest. In real experiments, with either quantum dots, atoms or some other physical system it may in practice be challenging to achieve degeneracy (or near degeneracy) between the excited states, meaning that in many practical cases the effects discussed will be seen only in a diluted or messier form.

We also considered a system with multiple ground states, finding an interesting arrangement in which a QE can be use to perform a two mode parity measurement on the WG modes.


\vspace{1pc}
Data Availability : The python codes used to perform the calculations and produce the plots in this paper are available at \cite{GreenScat_github_repo}.

\appendix
\section{Scattering Derivation}
\label{appendix_a}

We will first consider the situation where, initially, a single photon is propagating in the forwards direction of the WG and the QE is in one of its ground states. For completeness we will present the full derivation here, although many steps are the same as in the derivation for a two-level system \cite{Lang_perfect_2022}.

To find the scattering equation we will assume the QE and light field are initially uncoupled (a situation described with a non-interacting Hamiltonian $\hat{H}_0$) and the coupling between them ($\hat{H}_I$) will be added perturbatively, using an approach similar to that in \cite{2photon_emision_plasmon}.

The initial state of the system is described by a superposition of states selected from the spectrum of the non-interacting Hamiltonian with complex amplitudes $\gamma_{n}^{(0)}$:

\begin{equation}
\ket{\psi_{\text{Initial}}} = \sum_{n} \gamma_{n}^{(0)} \ket{n}.
\label{zeroth_order}
\end{equation}

\noindent
Perturbation theory can be used to calculate the state of the system at later times, accounting for the action of the interacting part of the Hamiltonian. The probability amplitude of each state at a later time is given by \cite{Grynberg_textbook}:

\begin{equation}
\gamma_{k}(t) = \sum_{m=0}^{\infty} \gamma_{k}^{(m)}(t),
\label{gam_sum}
\end{equation}

\noindent
where the $\gamma_{k}^{(m)}$'s are given by:

\begin{equation}
i \hbar \frac{d}{dt}\gamma_{k}^{(z+1)}(t) = \sum_n \bra{k} \hat{H}_{I} \ket{n} e^{i(E_k - E_n) t /\hbar} \gamma_{n}^{(z)}(t).
\label{perturber_diff}
\end{equation}

We take the initial state as:

\begin{equation}
\ket{\psi_{\text{Initial}}} = \ket{g_r 1_f},
\label{initial_cond}
\end{equation}

\noindent
i.e. one photon in the forwards waveguide mode and the QE in the $r^{\text{th}}$ ground state. The energy of this initial state is labelled $E_{\text{int}}$.

Overall the derivation will proceed as follows.  First, when the long photon approximation is applied the integral implied in equ.(\ref{perturber_diff}) can be replaced by a multiplying factor. Second, within the rotating wave approximation only even-numbered perturbation orders contribute. Together these simplifications allow all perturbation orders to be collected together into a single term using the geometric series identity. A choice of Hamiltonian is then inserted and some identities used to simplify. Finally we will assume the light is in a WG: which introduces a particular choice of photonic Green's function.

\subsection{Long Photon Limit}

We wish to evaluate the sum over all perturbation orders (equation \ref{gam_sum}) where each is calculated from the previous by equ.(\ref{perturber_diff}), which can be re-arranged:

\begin{equation}
\gamma_{k}^{(z+1)}(T) = \hspace{-1.4pc} \int \limits_{\text{Initial time}}^{T} \hspace{-0.6pc} \sum_n \bra{k} \hat{H}_{I} \ket{n}\frac{e^{i(E_k - E_n) t /\hbar} \gamma_{n}^{(z)}(t)}{i \hbar} dt.
\label{perturber_int}
\end{equation}

In this subsection it is shown that the time integral can be replaced by a antiderivative within the long photon limit. In this limit the input photon is infinitely long in time (i.e. single frequency).

To introduce this limit we impose the initial condition, equ.(\ref{initial_cond}) at $t=-\infty$ (instead of the more obvious $t=0$). We then require that the interaction Hamiltonian very slowly ``turns off'' at times approaching $\pm \infty$ as is done in \cite{Shen_Fan_Perturb, Loudon_textbook}. To achieve this one uses a modified Hamiltonian $\hat{H}_{I}'(t) = \hat{H}_{I}e^{-q|t|}$, where $q$ is some positive infinitesimal that is taken, in the limit, to zero. It works out that:

\begin{equation}
\begin{aligned}
\lim_{q\to0}\int_{-\infty}^{T} e^{i(a-b)t-q|t|} dt =& \frac{e^{i(a-b)T}}{i(a-b)} \\
=& \text{antiderivative}\left[e^{i(a-b)t}\right],
\end{aligned}
\label{anti_derivative_rule}
\end{equation}

\noindent
i.e. setting the lower limit to $-\infty$ and including this infinitesimal $q$ produces the anti-derivative with no $+C$ terms. Any $+C$ terms would be extremely problematic as by the $n^{th}$ integral there would be $n$ different such terms (the first of which would itself have been integrated $n-1$ times).

Physically the introduction of this infinitely slow turning off should make no difference to the result. At times $\pm\infty$ no part of the photon pulse (or at least only an infinitesimal part of its total amplitude) will be anywhere near the QE, so it cannot ``notice'' that the interaction has been turned off.


\subsection{Even orders}

The rotating-wave form of the interaction Hamiltonian we use (see later, equ. \ref{Rotating_Hamiltonian}) does one of two things each time it is applied. Either the QE is lifted up one of its optical transitions with the annihilation of a photon or the QE is lowered down a transition with the creation of a photon.

The QE is assumed to begin the scattering process in a ground state and after a sufficiently long time the QE must eventually decay back to a ground state. 

We now make use of the assumption that the QE is non-cascaded, meaning that excited states can only transition to ground states and ground states only to excited states. With a non-cascaded energy level arrangement the excitation number is conserved in the rotating wave approximation.

The application of the first Hamiltonian ($z=1$) will lift the QE from the ground state manifold to the excited state manifold (destroying the photon). A second application will return the QE to the ground state manifold (creating a photon) and so on. Only chains of an even length can contribute to the scattering. (In \cite{Feyn_Green} these perturbation orders are represented by Feynman diagrams; the relevant ones all posses an even number of vertices.)

The combinatorics of these chains of terms (for $z>0$) can be collected together as matrix products as:

\begin{widetext}
\begin{equation}
\gamma_{k}^{(2z)}(\infty) = \int_{-\infty}^{+\infty} \bra{k} \hat{H}_{I} \begin{pmatrix} \ket{e_1} & \ket{e_2} &\cdots & \ket{e_m} \end{pmatrix}  \left( \overline{\Delta}^{-1} \overline{\Gamma} \right)^{(z-1)}   \overline{\Delta}^{-1} \begin{pmatrix} \bra{e_1} \\ \bra{e_2} \\ \vdots \\ \bra{e_m} \end{pmatrix} \hat{H}_{I}\ket{g_r 1_f}  \frac{-e^{i(E_k - E_{\text{int}}) t /\hbar}}{i \hbar} dt,
\label{perturber_int_2}
\end{equation}
\end{widetext}

\noindent
where $\overline{\Gamma}$ is a $m$ by $m$ matrix, for $m$ the number of excited states. The $xy^{\text{th}}$ element of $\overline{\Gamma}$ is given by:

\begin{equation}
\overline{\Gamma}_{xy} = \bra{e_x} \hat{H}_{I} \left( \sum_{j} \sum_{n} \frac{\ket{g_n 1_j}\!\bra{g_n 1_j}}{(E_{gn} + \hbar \omega_j - E_{\text{int}})} \right) \hat{H}_{I} \ket{e_y}.
\label{V}
\end{equation}
\noindent
The sum over $j$ is a sum over all photon modes, the sum $n$ is over all QE ground states. The matrix $\overline{\Gamma}$ represents a sum over every way in which two applications of the Hamiltonian can move complex amplitude from excited state $y$ to each other excited state $x$.

The term $\overline{\Delta}$ is the diagonal detuning matrix, given by:

\begin{equation}
\overline{\Delta} = \begin{pmatrix}
	E_{e_1} - E_{\text{int}} & 0 & \cdots & 0 \\
	0 & E_{e_2} - E_{\text{int}} & \cdots & 0  \\
	\vdots  & \vdots   & \ddots &  \vdots \\
     0  &   0    &    \cdots   & E_{e_m} - E_{\text{int}} \\
\end{pmatrix}.
\vspace{0.25pc}
\label{detuning_matrix}
\end{equation}

Equation (\ref{perturber_int_2}) shows that each even perturbation order is given the previous one with an extra multiplying factor of $\overline{\Delta}^{-1} \overline{\Gamma}$. The geometric series identity for matrices states that (assuming that the sum converges):

\begin{equation}
\sum_{n=0}^{\infty} \overline{A}^{n} = \left(\mathbb{1} - \overline{A} \right)^{-1}.
\label{geom_mat}
\end{equation}

\noindent
Using this one can collect all perturbation orders together. The zeroth order must also be included as an additive Kronecker-Delta, equ.(\ref{zeroth_order}), and the integral is evaluated:

\begin{widetext}
\begin{equation}
\gamma_{k}(\infty) = \delta_{g_r 1_f}^{k}  + 2 \pi i \delta(E_k - E_{\text{int}})   \bra{k} \hat{H}_{I} \vec{\ket{e}} \left(\mathbb{1} - \overline{\Delta}^{-1} \overline{\Gamma} \right)^{-1} \overline{\Delta}^{-1}  \vec{\bra{e}} \hat{H}_{I} \ket{g_r 1_f}.
\label{summed_gam}
\end{equation}

With $\vec{\ket{e}}=  (\vec{\bra{e}})^{\dagger}  =  \begin{pmatrix} \ket{e_1} & \ket{e_2} &\cdots & \ket{e_m} \end{pmatrix}$.

The role of detuning is simplified by pre-multiplying inside the inverse and outside it (on the right) with the detuning matrix $\overline{\Delta}$,

\begin{equation}
\gamma_{k}(\infty) = \delta_{g_r 1_f}^{k} + 2 \pi i \delta(E_k - E_{\text{int}}) \bra{k} \hat{H}_{I} \vec{\ket{e}} \left(\overline{\Delta} - \overline{\Gamma} \right)^{-1} \vec{\bra{e}} \hat{H}_{I}\ket{g_r 1_f}.
\label{before_ham}
\end{equation}
\end{widetext}

The result to here is independent of the exact form the interaction Hamiltonian takes, so long as it describes a non-cascaded set of QE energy levels within the rotating wave approximation. 

Only a single photon is described by this theory, inclusion of multiple photons would expand the state space and hence greatly expand the number of paths.

The Hamiltonian describing light-matter coupling in a dielectric medium is inserted to solve for both $\overline{\Gamma}$. This proceeds as described in previous papers (eg. see supplement of \cite{Lang_perfect_2022}), although for completeness it is detailed in appendix \ref{appendix_Greens_Hamiltonian}.

The result is that the elements of $\overline{\Gamma}$ are given by

\begin{equation}
\overline{\Gamma}_{xy} = - \sum_{n} \big( \mathbf{d}_{nx} \cdot \textbf{G}^{*}(\mathbf{r}, \mathbf{r}, \omega_n) \cdot \mathbf{d}_{ny}^{*}  \big)/\epsilon_0,
\label{final_G2}
\end{equation}
where $\mathbf{d}_{nx}$ is the dipole associated with a transition from ground state $n$ to excited state $x$.

It is worth discussing the interpretation of these $\overline{\Gamma}$ terms, starting with the diagonal terms where $x=y$. The imaginary parts of these diagonal elements are often called the Local Density of States (LDOS) and measure the combined decay rate of excited state $x$ to all the ground states it couples to (in units of energy and up to a factor of 2) \cite{G_anisotropy}.

Similarly the real parts represent a modification to the energy of excited states due to there interaction with the electromagnetic environment. In this case the excited states receive additive energy shifts from each ground state they can decay into.

Now consider the off-diagonal terms. When two different excited states can both decay to one or more of the same ground states equ.(\ref{final_G2}) indicates that they interact with one another. In simplest case of only two initially degenerate excited states this interaction promotes energy shifts and lifetime changes in there symmetric and anti-symmetric combinations. For example the real part of an off-diagonal term indicates a shift in the resonant frequency of the symmetric state \cite{Green_superrad}, this is called a dipole-dipole shift \cite{Dipole_dipole_lamb_shift}. The imaginary part of an off-diagonal term instead controls superradiance/subradiance, that is, changes in the lifetime of the symmetric and anti-symmetric superposition states. For a simple example of why this term must exist consider a \emph{V}-type QE arrangement of two excited states that decay to a single ground state. If this QE is initially prepared in a superposition of the two excited states the radiation from the two decay channels will interfere - possibly destructively. In the destructive case the total outgoing photon amplitude is reduced, implying an extended lifetime.

Where the two excited states do not share a ground state (for example an \emph{II} arrangement) interference of this type is not possible as the photon amplitude from each transition is entangled with the ground state (eg. $\ket{g_1 1_j} - \ket{g_1 1_j} = 0$ but $\ket{g_1 1_j} - \ket{g_2 1_j} \neq 0$). This is reflected in the shared $n$ subscripts on the dipoles in equation (\ref{final_G2}).

Note that the energy shifts are given by the real part of the whole term, including both dipoles and Green's function (and the broadening by the imaginary part), as in, for example, \cite{Liu_real_im_mixing_2018}. Variants like: $\text{Decay rate} \propto \mathbf{d}_{nx} \cdot \text{Im}[\textbf{G}] \cdot \mathbf{d}_{ny}^{*}$ are commonly equivalent, for example when only real dipoles are considered or when the dipoles either side of the Green's function are identical, meaning that variants like this (placing the dipoles outside the Im or Re operations), are sometimes encountered in the literature \cite{Quantum_Greens, Dipole_dipole_lamb_shift, Welsh_dipole_dipole}. In the general case the real and imaginary parts of the entire expression, including the dipoles, should be taken, matching, for example, \cite{}.

Next the $\vec{\bra{e}} \hat{H}_{I} \ket{g_r 1_f}$ terms are evaluated. They are given by \cite{Hughes_laser_review, denmark, Hughes_multidot_entanglement}:

\begin{equation}
\bra{e_y} \hat{H}_{I} \ket{g_r 1_f} \propto \sqrt{\frac{\omega \hbar}{2\epsilon_0}} \mathbf{d}_{ry}^* \cdot  \mathbf{E}_{f}(\mathbf{r}).
\label{kets_to_dot_product}
\end{equation}

\noindent 
With $\mathbf{E}_{f}$ the electric field of the mode $f$, and $\omega$ the angular frequency of a transition. The constant of proportionality depends on how the modes are indexed and takes different values for modes distributed in wave-vector, frequency and energy \cite{denmark}.

We assume that the scattered photon is detected by some device that has a finite energy resolution - represented by integrating the overall expression, equ.(\ref{before_ham}), across a small interval of energies that are not resolvable from one another. One could instead integrate over frequency, or wave vector, producing the same result but with a different prefactor.

Any consistent choice, where the modes in equ.(\ref{kets_to_dot_product}) are indexed by the same parameter that is integrated over, produces the same result as any other consistent choice. As a result the total intensity radiated into the spectral window is independent of the modes used to represent the system (but of course the intensity into any given mode critically depends on the definition of the modes one is using).

Inserting this relation for the coupling (\ref{kets_to_dot_product}), re-arranging and taking this integral over a tiny spectral window is equivalent to the following substitutions to equ.(\ref{before_ham}):

\begin{equation}
\begin{split}
2 \pi i \delta(E_{k} - E_{\text{int}}) &\rightarrow \frac{i z }{\epsilon_0} \\
\vec{\bra{e}} \hat{H}_{I} \ket{g_r 1_f}  &\rightarrow \begin{pmatrix} \mathbf{d}_{r1}^* & \mathbf{d}_{r2}^* &\cdots & \mathbf{d}_{rm}^* \end{pmatrix}^{\text{T}} \cdot  \mathbf{E}_{f}(\mathbf{r}) \\
\bra{g_k 1_k} \hat{H}_{I} \vec{\ket{e}} &\rightarrow \begin{pmatrix} \mathbf{d}_{k1} & \mathbf{d}_{k2} &\cdots & \mathbf{d}_{km} \end{pmatrix} \cdot  \mathbf{E}_{k}^*(\mathbf{r}) 
\label{subs}
\end{split}
\end{equation}

\noindent
where the $\mathbf{E}$ fields are given for the Bloch parts of modes indexed discretely by their $k$-vector. The $\delta$-function ensures energy conservation, so that $\omega_{k} = \omega_{f} +(E_{g_r} - E_{g_k})/\hbar$. The constant $z$ is a scaling factor that accounts for the density of states. In a photonic crystal waveguide $z = a \omega/(2 |v_g|)$ where $v_g$ is the group velocity (assumed the same at all relevant frequencies) $a$ is the lattice constant, and $\omega$ is the angular frequency of the photon, where we assume the input and output photons are close enough in frequency that a single frequency variable for all transitions is sufficient.

Up to here the derivation is general to any photonic environment (except for the value taken by $z$). The only remaining step is to introduce a specific photonic environment by choosing a Green's function. Inserting (\ref{Greens_function}) this into the equation for $\overline{\Gamma}$, equ.(\ref{final_v}), inserting that and equ.(\ref{subs}) into equ.(\ref{before_ham}) and re-arranging:

\begin{equation}
\begin{split}
\gamma_{k}(\infty) &= \delta_{g_r 1_f}^{k} \\
 & - \overrightarrow{\mathbf{d}_{k:}}\cdot  \mathbf{E}_{k}^*(\mathbf{r}) \Big( \overline{X}  + \frac{\overline{L}}{z}  + \frac{i\epsilon_0 \overline{\Delta}}{z} \Big)^{-1}
\overrightarrow{\mathbf{d}_{r:}^{*}}^{\text{T}} \cdot  \mathbf{E}_{f}(\mathbf{r}),
\label{final}
\end{split}
\end{equation}

\noindent
where:

\begin{equation}
\begin{split}
\overline{X}_{xy} &= \sum_n \frac{1}{2} \mathbf{d}_{nx} \cdot \left( \mathbf{E}_{f}^{*} \mathbf{E}_{f}  + \mathbf{E}_{b}^{*} \mathbf{E}_{b} \right) \cdot \mathbf{d}_{ny}^{*},\\
\overline{L}_{xy} &= \sum_n \mathbf{d}_{nx} \cdot \textbf{G}_{\text{loss}}^{*} \cdot \mathbf{d}_{ny}^{*},
\end{split}
\end{equation}

\noindent
and $\overrightarrow{\mathbf{d}_{k:}} =  \begin{pmatrix} \mathbf{d}_{k1} & \mathbf{d}_{k2} &\cdots & \mathbf{d}_{km} \end{pmatrix}$. It has been assumed that the $\mathbf{E}$s are independent of frequency.


\section{Emission Derivation}

The previous section considered the situation in which a single photon of narrow frequency scatters from the multilevel system. Another relevant situation is that where the system is initially prepared in the excited state manifold and decays into the ground state manifold.

A derivation is also presented for this case, both for completeness and to have both using the same notation. Comparable derivations, with differences in assumptions or the circumstances considered are used as a basis 
\cite{stephen_hughes_mollow_triplet_metalic_np, G_anisotropy, dipole_dipole_interactions_near_fibre, Welsh_dipole_dipole, Hughes_double_dot_mollow}.

In summary, the process follows a fairly standard opening for deriving a Master equation for an open quantum system, by first moving to the interaction picture then substituting the Shr{\"o}dinger equation into itself, producing an equation at second order in the Hamiltonian for the rate of change of the system state which has Born and Markov approximations applied to it. As was seen for scattering, there exists a significant simplification of the square of the interaction Hamiltonian, hence second order offers a simplification.

It is assumed that initially the electromagnetic fields are in a pure vacuum state.

Starting with the Shr{\"o}dinger equation:

\begin{equation}
\dot{\rho}(t) = \frac{ - \im}{\hbar} [\hat{H}(t), \rho(t)] 
\end{equation}

\begin{equation}
\rho(t) = \rho(0) + \int_0^t \frac{ - \im}{\hbar} [\hat{H}(t'), \rho(t')] dt' 
\end{equation}

and subbing it into itself;

\begin{equation}
\dot{\rho}(t) = \frac{ - \im}{\hbar} [\hat{H}(t), \rho(0)] - \int_0^t \frac{ - 1}{\hbar^2} [H(t), [\hat{H}(t'), \rho(t')]] dt' \,.
\end{equation}

We introduce a new parameter, $\tau = t-t'$. If $t$ is interpreted as the present time then $\tau$ is how long ago something happened.

\begin{equation}
\dot{\rho}(t) = \frac{ - \im}{\hbar} [\hat{H}(t), \rho(0)] - \int_{t}^0 \frac{ - 1}{\hbar^2} [\hat{H}(t), [\hat{H}(t-\tau), \rho(t-\tau)]] d\tau
\label{Double_H}
\end{equation}

\subsection{Interaction Picture}

Moving to the interaction picture, the only part of the Hamiltonian remaining is the interaction part. However, it is transformed by the move to the interaction picture: each input whirs in a positive sense at the lab-frame frequency of the corresponding state, and each output whirs in the negative sense at the frequency of that output state

\begin{equation}
\hat{H}_{II}(t) =  \sum_{\psi, \psi'} \ket{\psi}\!\bra{\psi} \hat{H}_I \ket{\psi'}\!\bra{\psi'} \exp(\im t (\omega_{\psi} - \omega_{\psi'})).
\end{equation}
where the sum is over the eigenstates of the non-interacting lab frame Hamiltonian
\begin{equation}
\hat{H}_0 \ket{\psi} = \hbar \omega_{\psi} \ket{\psi}.
\end{equation}

The average over time of $\exp(\im \omega t)$ is zero, and consequently all Hamiltonian terms (except those where $\omega_\psi = \omega_{\psi'}$) average to zero over a sufficiently long timescale. Terms connecting states that are distant in energy will spin rapidly and average to zero even over short times, while those close in energy will only average zero over longer times. The relevant timescale for comparison is that implied by the elements of $|\hat{H}_{II}|$, as this is how long the Hamiltonian would take to do anything, if it were not constantly changing phase (due to the exponentials). For simplicity in this case, we will assume this timescale is very long (Born approximation \cite{Carmichael_book}).

As we restrict the analysis to the single-excitation subspace a density matrix can be expressed using a combination of terms referring to: (i) an excited state with no photon ($\ket{e_n, 0}\bra{e_m,0}$), or (ii) a ground state with one photon in some mode $j$ or $j'$ ($\ket{g_n, 1_j}\bra{g_m, 1_{j'}}$. Coherences between these states occur when $n\neq m$ and $j \neq j'$. We also have cross-type coherneces of the form $\ket{e_n, 0}\bra{g_m,j'}$ and $\ket{g_n, j, 0}\bra{e_m}$.

Tracing over the optical degrees of freedom leads to an equation for the motion of $\hat{\rho}_{a}$, given by the partial trace of $\hat{\rho}$. $\hat{\rho}_{a}$ spans the states $\ket{e_n}$ and $\ket{g_n}$. Note that, as in the full system, every instance of an excited state goes with vacuum and every instance of a ground state goes with a photon, the reduced density matrix cannot have coherences between the excited and ground state manifolds. (However, coherences within either the excited state or ground state manifold are possible).

With the assumption that there is only one excitation, the partial trace can be expressed:

\begin{equation}
\begin{split}
\hat{\rho}_a &= \sum_{n,m} \ket{e_n}\!\bra{e_n, 0} \hat{\rho} \ket{e_m,0}\!\bra{e_m} \\ &+ \sum_{n,m, j} \ket{g_n}\!\bra{g_n, 1_j} \hat{\rho} \ket{g_m, 1_j}\!\bra{g_m}.
\end{split}
\end{equation}
(There is only one $j$ variable describing the photon states as we are tracing over them, so should be diagonal on them. Unlike the excited and ground states, which are in the part of the space being kept, so we have two variables to include off-diagonals.)

The interaction Hamiltonian can be expressed $\hat{H}_{I} = \sum_{egp} \hat{T}_{egp} + \text{H.C.}$, with each $\hat{T}_{egp}$ an operator representing an absorption of a photon in state $p$, while moving the system from ground state $g$ to excited state $e$. The explicit Hamiltonain (where these $\hat{T}$ operators are detailed) is given in (equ. \ref{Rotating_Hamiltonian}).

This gives the interaction picture Hamiltonain:

\begin{equation}
\hat{H}_{II}(t) = \sum_{egp} \hat{T}_{egp} \exp(\im t (E_e - E_g - \omega_p)) + \text{H.C.}\,.
\end{equation}

Substituting this into (equ.\ref{Double_H}), assuming an initial state in the excited state manifold with zero photons, and taking the partial trace, we first see that the leading term (with one copy of the Hamiltonain) vanishes. The remaining part with the double commutator contains four components, resembling $\hat{H}(t) \hat{H}(t-\tau) \rho$, $\rho\hat{H}(t-\tau) \hat{H}(t)$,  $\hat{H}(t-\tau) \rho \hat{H}(t)$ and  $\hat{H}(t) \rho \hat{H}(t-\tau)$. Given the form of $\hat{H}_{II}$, the initial excited state and the partial trace these simplify. For example consider the $HH\rho$ part.

\begin{widetext}
\begin{equation}
HH\rho = - \sum_{egp\,e'g'p'} \int_0^{t} \frac{1}{\hbar^2} \hat{T}_{egp} \hat{T}^{\dagger}_{e'g'p'} \exp\left( \frac{\im t}{\hbar}(E_e - E_g - \hbar\omega_p - E_{e'} + E_{g'} + \hbar\omega_{p'})\right) \exp\left(\frac{\im \tau}{\hbar} (- E_{e'} + E_{g'} + \hbar\omega_{p'})\right) \rho(t-\tau) d\tau
\label{HHp}
\end{equation}
\end{widetext}
Where for brevity, the (somewhat ceremonial) $\ket{e}\!\bra{e, 0}$ terms sandwhiching the expression have been omitted.

Consider the part of this expression of the form:

\begin{equation}
\int_0^t  \exp \left( \frac{\im \tau}{\hbar} (- E_{e'} + E_{g'} + \hbar \omega_{p'}) \right) \, \rho(t-\tau) \, d\tau \,
\end{equation}
This part is simplified by making two assumptions. First, the $t$ integration limit is replaced with $\infty$. Second, an exponentially decaying envelope function is introduced, and the limit of this function decaying infinitely slowly is taken (so, the limit of it doing nothing), as in (equ. \ref{anti_derivative_rule}). We also ignore the $\tau$ dependence in $\rho(t-\tau)$.

These assumptions collectively encapsulate a Markov assumption \cite{Carmichael_book}, and lead to this integral being replaced with:

\begin{equation}
\frac{\hbar \, \rho(t)}{\im (E_{e'} - E_{g'} - \hbar \omega_{p'})}\,.
\end{equation}

We also use the fact that the $\hat{T}_{egp} \hat{T}^{\dagger}_{e'g'p'}$ term is zero if $g\neq g'$ or $p\neq p'$, which leads to

\begin{equation}
HH\rho = -\sum_{xy} \ket{e_x}\frac{1}{\im \hbar} \overline{\Gamma}_{xy} \exp\left( \frac{\im t}{\hbar}(E_{ex} - E_{ey})\right) \bra{e_y} \rho_a
\end{equation}
with $\overline{\Gamma}$ from (equ.\ref{V}), with $E_{\text{Int}}$ understood to still be the initial energy, so in this case $E_{ey}$.

The RHS defines an operator acting on $\rho_a$, which can be transformed back out of the interaction picture to remove the time dependence.

This was the first of the four terms from the double commutator. The term of shape $\rho H H$ is found analogously, and is just a flipped version of the term just discussed, with an extra $-1$ multiplying it.

The sandwich terms are slightly different, so for completeness are considered explicitly.

In the sandwhich terms, we are taking the partial trace over a $\rho$ with a $\hat{T}$ or $\hat{T}^{\dagger}$ term on either side of it. The partial trace over the photon degrees of freedom introduces a requirement that $p=p'$, similar to what emerged with the adjacent $\hat{T}$ terms previously. This is relates to the cyclic property of the trace operation. However, its only a partial trace, so the parts relating to the QE are left alone, meaning there is no requirement for $g=g'$.

Subbing it in, and resolving the $\tau$ integral we find (where the RHS is not literal, but is referring to the term of that shape):

\begin{widetext}
\begin{equation}
H(t) \rho {H}(t-\tau) = \sum_{egp e'g'} \ket{g}\!\bra{g, 1_p}  \hat{T}_{egp} \ket{e,0}\!\bra{e,0} \rho \ket{e',0}\!\bra{e',0} \hat{T}^{\dagger}_{e'g'p'} \ket{g', 1_p}\!\bra{g'} \frac{\im \hbar \exp\left( \frac{\im t}{\hbar}
(E_e - E_g + E_{g'} - E_{e'})
\right)}{(E_{g'} + \hbar \omega_p - E_{e'})} 
\end{equation}
\end{widetext}

Defining $\overline{W}$:

\begin{equation}
\overline{W}_{xynm} = \bra{e_x} \hat{H}_{I} \left( \sum_{j} \frac{\ket{g_n 1_j}\!\bra{g_m 1_j}}{(E_{gm} + \hbar \omega_j - E_{ey})} \right) \hat{H}_{I} \ket{e_y}.
\label{W}
\end{equation}
(with the consequence of this is that $\overline{\Gamma}_{xy} = \sum_n \overline{W}_{xynn}$, note the repeated $n$ index), it can be simplified:

\begin{equation}
\begin{split}
H(t) \rho {H}(t-\tau)) = \sum_{xynm} \ket{g_n}\!\bra{e_x} \rho_a \ket{e_y}\!\bra{g_m} \overline{W}_{xynm} \\
\times \exp\left( \frac{\im t}{\hbar}(E_{ex} - E_{gn} + E_{gm} - E_{ey})\right) \im \hbar
\end{split}
\end{equation}

The other sandwich term, $H(t-\tau) \rho H(t)$ is similar. It differs in that it contains a $\overline{W}_{yxmn}^{*}$ (ie indices swapped, and complex conjugate) and an additional factor of $-1$.

Transforming back out of the rotating frame, we can re-arrange the overall expression to read:

\begin{equation}
\begin{split}
\frac{d\hat{\rho}_a(t)}{dt} =& \sum_{xy} \frac{\im}{\hbar} \left( \overline{\Gamma}_{xy} \ket{e_x}\!\bra{e_y} \hat{\rho}_a - \overline{\Gamma}^*_{xy} \hat{\rho}_a \ket{e_x}\!\bra{e_y}
\right) \\
&- \frac{\im}{\hbar} [\hat{H_0}, \rho_a(t)] \\
&-\frac{\im}{\hbar} \sum_{xynm} \overline{V}_{xynm} \ket{g_n}\!\bra{e_x} \rho_a \ket{e_y}\!\bra{g_m} \,.
\end{split}
\end{equation}
With
\begin{equation}
\overline{V}_{xynm} = \overline{W}_{xynm} -  \overline{W}_{yxmn}^{*}\,.
\end{equation}
The $\hat{H}_0$ term comes from the fact we transformed back out of the rotating frame.

Note that the first term can written in the form of a commutator with the real part of $\overline{\Gamma}$ and an anti-comutator with the imaginary part of it, as in for example \cite{G_anisotropy}.

\section{Green's function Hamiltonian}
\label{appendix_Greens_Hamiltonian}

We choose a Hamiltonian that describes the light-matter interaction in terms of the photonic Green's function. The reason for this choice is that one of our primary interests is the polarisation degree of freedom and its connection to the propagation direction, a connection which is well modelled by Green's functions without imposing it by hand \cite{Young_PRL}.

Within the rotating wave approximation the interaction Hamiltonian for a single dipole transition (from ground state $x$ to excited state $y$) is given by \cite{stephen_hughes_mollow_triplet_metalic_np, welsch2}:

\begin{widetext}
\begin{equation}
\hat{H}_{xy} = - i \hat{\sigma}_{xy}^{+} \mathbf{d}_{xy} \cdot   \iint \textbf{G}(\mathbf{r}, \mathbf{r}', \omega') \sqrt{ \frac{\hbar \text{Im}[\epsilon(\mathbf{r}',\omega')]}{\epsilon_{0} \pi}} \mathbf{\hat{f}}(\mathbf{r}',\omega') d^{3}\mathbf{r}' d\omega' - \text{H.C.},
\label{Rotating_Hamiltonian}
\end{equation}
\end{widetext}

\noindent where $\hat{\sigma}_{xy}^{+}$ is the raising operator from ground state $x$ to excited state $y$. H.C. indicates the Hermitian Conjugate of the 1$^{\text{st}}$ term, this H.C. part represents photon emission with the decay of the QE to ground state $x$ from excited state $y$.

The $\mathbf{\hat{f}}$/$\mathbf{\hat{f}}^{\dagger}$ are bosonic field annihilation/creation operators that represent basic excitations of the dielectic material degrees of freedom and the related electromagnetic field \cite{green_bible, f_operators_momentum}. $\epsilon$ is the relative electric permittivity, its imaginary part corresponds to the damping (loss) of electromagnetic radiation in the material.

A QE of $m$ excited states and $n$ ground states will have a total of $n\times m$ transitions, each represented by a copy of equ.(\ref{Rotating_Hamiltonian}) with the appropriate choice of dipole. Forbidden transitions can be  modelled by setting $|\mathbf{d}| = 0$. The total interaction Hamiltonian is given by the sum of all these single-transition terms.

This Hamiltonian is used to evaluate the elements of $\overline{\Gamma}$. This appendix describes the calculation of these terms.

Beginning with $\overline{\Gamma}$ we first throw away all the parts of the Hamiltonian that are not selected for by the raising/lowering operators.

\begin{equation}
\overline{\Gamma}_{xy} = \bra{e_x}  \left( \sum_{j} \sum_{n} \hat{H}_{nx} \frac{\ket{g_n 1_j} \! \bra{g_n 1_j}}{(E_{g_n 1_j} - E_{\text{int}})}\hat{H}_{ny}  \right) \ket{e_y}.
\end{equation}

The sum over all photon states ($\sum_{j}$) can be carried out in the position basis, with each photon state represented by a unique combination of position $\mathbf{R}$, frequency $\omega$ and one of three orthogonal polarisations $p$. Position and frequency are continuous so the sum over these parameters is replaced with an integral. Overall $\sum_j \rightarrow \sum_{p=1}^{3} \iint d^{3}\mathbf{R} d\omega$ and $\ket{1_j} \rightarrow \ket{1_{\mathbf{R},\omega,p}}$.

The field operators obey the commutation relation, $[\hat{f}_{a}(\mathbf{r},\omega), \hat{f}^{\dagger}_{b}(\mathbf{r}',\omega')] = \delta(\mathbf{r}-\mathbf{r}') \delta(\omega-\omega') \delta_{a}^{b}$, where $a$,$b$ denote the vector components of the vector-operators \cite{welsch1}. With $\ket{1_{\mathbf{R},\omega,p}} = \hat{f}^{\dagger}_{p}(\mathbf{R},\omega)\ket{0}$ this implies that

\begin{equation}
\sum_{p=1}^{3} \bra{g_n 1_{\mathbf{R},\omega,\mathbf{p}}} \hat{\sigma}_{ny}^{-} \mathbf{\hat{f}}^{\dagger}(\mathbf{r}'',\omega'') \ket{e_y} = \delta(\mathbf{R}-\mathbf{r}'') \delta(\omega-\omega'') \mathbf{I},
\end{equation}

\noindent
where $\mathbf{I}$ is the identity tensor (acting on polarisation). Inserting these relations one finds that the innermost integrals (those from the Hamiltonians) are evaluated by these $\delta$-functions:

\begin{equation}
\begin{split}
\overline{\Gamma}_{xy} =& - \iint \sum_{n}  \frac{\mathbf{d}_{nx} \cdot \textbf{G}(\mathbf{r}, \mathbf{R}, \omega) \cdot \mathbf{d}_{ny}^{*} \cdot \mathbf{G}^{*}(\mathbf{r}, \mathbf{R}, \omega)}{\epsilon_0 \pi (E_{g1_j} - E_{\text{int}})}\\
&\times \hbar \text{Im}[\epsilon(\mathbf{R},\omega)] d^{3}\mathbf{R} d\omega.
\end{split}
\end{equation}

The Green's functions identities \cite{stephen_hughes_mollow_triplet_metalic_np}:

\begin{equation}
\begin{split}
\mathbf{G}(\mathbf{r},\mathbf{R},\omega) &= \mathbf{G}^{T}(\mathbf{R},\mathbf{r},\omega),\\
\mathbf{d}^* \cdot  \mathbf{G}^{*T} &= \mathbf{G}^{*} \cdot  \mathbf{d}^*,\\
\int \text{Im}[\epsilon(\mathbf{R},\omega)] \mathbf{G}(\mathbf{r},\mathbf{R},\omega) \mathbf{G}^{*}(\mathbf{R},\mathbf{r}',\omega)d^3\mathbf{R} &= \text{Im}[\mathbf{G(\mathbf{r},\mathbf{r}',\omega)}],
\end{split}
\end{equation}

\noindent
can be exploited to simplify to:

\begin{equation}
\overline{\Gamma}_{xy} =- \int \sum_{n}  \frac{\hbar \hspace{0.3pc} \mathbf{d}_{nx} \cdot \text{Im}[\textbf{G}(\mathbf{r}, \mathbf{r}, \omega)] \cdot \mathbf{d}_{ny}^{*}}{\epsilon_0 \pi (E_{g_n 1_j} - E_{\text{int}})} d\omega.
\end{equation}

The substitutions $E_{g_n 1_j} = E_{g_n} + E_{1_j}$, $\omega_n = (E_{\text{int}} - E_{g_n})/\hbar$ and $\omega = E_{1_j}/\hbar$ are made and we exploit the relation that, under an integral, $1/(\omega - \omega_n) = -i \pi \delta(\omega - \omega_n) + \mathcal{P} /(\omega - \omega_n)$ with $\mathcal{P}$ the principal value \cite{SH_P_value, dipole_dipole_interactions_near_fibre}:

\begin{equation}
\begin{split}
\overline{\Gamma}_{xy} = \sum_{n}  &i \frac{\mathbf{d}_{nx} \cdot \text{Im}[\textbf{G}(\mathbf{r}, \mathbf{r}, \omega_n)] \cdot \mathbf{d}_{ny}^{*}}{\epsilon_0} \\
- &\int_0^{\infty}  \mathcal{P} \frac{ \mathbf{d}_{nx} \cdot \text{Im}[\textbf{G}(\mathbf{r}, \mathbf{r}, \omega)] \cdot \mathbf{d}_{ny}^{*}}{\epsilon_0 \pi (\omega - \omega_n)} d\omega.
\end{split}
\end{equation}

The integral in the second term arose from the need to integrate over photons of all frequencies. We will make the approximation that the integral can be taken from $-\infty \rightarrow \infty$ instead of $0\rightarrow \infty$, which offers a significant simplification by using the Kramers–Kronig relation. This is an approximation, and this extended integration range amounts to neglecting an additional integral term discussed in detail in \cite{Dipole_dipole_lamb_shift}, which is only occasionally relevant. Making the approximation produces:

\begin{equation}
\begin{split}
\overline{\Gamma}_{xy} = \sum_{n} \big( i &\mathbf{d}_{nx} \cdot \text{Im}[\textbf{G}(\mathbf{r}, \mathbf{r}, \omega_n)] \cdot \mathbf{d}_{ny}^{*} \\- &\mathbf{d}_{nx} \cdot \text{Re}[\textbf{G}(\mathbf{r}, \mathbf{r}, \omega_n)] \cdot \mathbf{d}_{ny}^{*} \big)/\epsilon_0
\end{split}
\label{final_v}
\end{equation}

\begin{equation}
= - \sum_{n} \mathbf{d}_{nx} \cdot \textbf{G}^{*}(\mathbf{r}, \mathbf{r}, \omega_n) \cdot \mathbf{d}_{ny}^{*} /\epsilon_0.
\end{equation}

From following the same steps:

\begin{equation}
\overline{W}_{xynm} = - \mathbf{d}_{nx} \cdot \textbf{G}^{*}(\mathbf{r}, \mathbf{r}, \omega_m) \cdot \mathbf{d}_{my}^{*} /\epsilon_0.
\end{equation}

The WG Green's function (equation \ref{Greens_function}), is linearly proportional to frequency. The different transitions in the QE may correspond to photons of unequal frequencies and consequently see slightly different Green's functions. We will not include this effect, instead assuming all transitions are relatively close in close frequency, a common assumption \cite{V_level_Zoller_2016}. This assumption is also made in the relation \ref{kets_to_dot_product} used previously, using a single frequency for all these terms.

When the QE decays via photon emission, it is possible for the model to track which mode the photon was radiated into. The decay terms ($\overline{W}$) are linear in the Green's function, which can be split into a sum of Green's function terms for the different modes, for example as $\textbf{G} = \textbf{G}_f + \textbf{G}_b + \textbf{G}_{\text{loss}}$ for the forward, backward and loss parts of the Green's function (see equation \ref{Greens_function}). This allows $\overline{W}$ to be broken up into three parts, one using each part of the Green's function, and we know that a decay to the ground state manifold mediated by the $\overline{W}$ term related to $\textbf{G}_f$ means a photon radiated into the forward mode.

\bibliography{bibliogrpahy}

\end{document}